\documentclass[twocolumn]{aa}
\usepackage{graphicx}
\usepackage{txfonts}
\begin{document}
   \title{Polarimetric evolution of V838 Monocerotis
          \thanks{Based on observations collected at Padua Astronomical
                  Observatory at Asiago and Crimean Astrophysical Observatory}} 


   \author{S. Desidera
          \inst{1},
           E. Giro
           \inst{1},
           U. Munari
          \inst{2},
           Y.S. Efimov 
          \inst{3},
           A. Henden
          \inst{4},
           S. Benetti
          \inst{1},
           T. Tomov
          \inst{5},
           A. Bianchini
          \inst{6}
          \and
           C. Pernechele
          \inst{1}
}

   \authorrunning{S. Desidera et al.}

   \offprints{S. Desidera}

   \institute{INAF -- Osservatorio Astronomico di Padova, 
              Vicolo dell' Osservatorio 5, I-35122, Padova, Italy \\
              \email{desidera,giro,benetti,pernechele@pd.astro.it}         
              \and 
              INAF -- Osservatorio Astronomico di Padova - Sede di Asiago, 
              I-36012, Asiago, Italy \\
              \email{munari@pd.astro.it}         
          \and
          Crimean Astrophysical Observatory and Isaac Newton Institute of 
          Chile, Crimean Branch, Ukraine \\
           \email{efimov@crao.crimea.ua}
          \and
           Universities Space Research Association/U. S. Naval Observatory
           Flagstaff Station, P. O. Box 1149, Flagstaff AZ 86002-1149, USA  \\
           \email{aah@nofs.navy.mil}
          \and
         Centre for Astronomy Nicholaus Copernicus University,
          ul. Gagarina 11, 87-100 Torun, Poland \\
          \email{Toma.Tomov@astri.uni.torun.pl}
          \and
         Universit\'a di Padova, Dipartimento di Astronomia,
         Vicolo dell'Osservatorio 2, I-35122, Padova, Italy \\
         \email{bianchini@pd.astro.it}
}

   \date{}

   \abstract{We present the results of our polarimetric and 
             spectropolarimetric 
             monitoring of V838 Monocerotis, performed  at Asiago
             and Crimean  observatories during and after the multiple outbursts
             that occurred in January-March 2002.
             The polarization of the object is
             mainly due to interstellar polarization ($P\sim 2.48$\%).
             Intrinsic polarization up to $\sim 0.7$\% at 5000~\AA~ 
             is present during
             the second maximum of the object (February 2002). This 
             intrinsic component increases toward shorter wavelengths
             but our limited spectral coverage (5000-7500~\AA) does not allow
             conclusive inferences about its origin. 
             A strong depolarization across the $H_{\alpha}$ profile
             is observed. 
             The interstellar polarization gives a  lower limit
             to the  reddening of {\em E(B-V)} $>0.28$, with
             {\em E(B-V)} $\sim$ 0.5 being the most probable value.
             A normal ratio of total to selective absorption 
             ($R_{V}=3.22\pm0.17$) was derived
             from the wavelength of maximum interstellar polarization.
             This suggests a low (if any) contribution by
             circumstellar material with peculiar dust to gas ratio.
             A polarimetric map of a portion of the light echo
             shows a complex polarization distribution reaching
             $P_{max}=45$\%.

   }

   \maketitle
%

\section{Introduction}

V838 Monocerotis developed a spectacular multiple outburst
in January-March 2002, reaching V=6.7. Its spectral
characteristics changed dramatically during its evolution.
The progenitor had the temperature of a F star, while during
the outburst V838 Mon evolved from a cool K giant to a late M
giant. Profile of spectral lines also changed during
the ourburst, and a strong $H_{\alpha}$ emission appeared
during the second maximum, when the object reached its peak
visual magnitude. A prominent light echo was discovered
by Henden et al.~(\cite{henden02}).
The evolution of V838 Mon from January to April 2002 is 
described by Munari et al.~(\cite{munari02}). Further
photometric and spectroscopic
observations were presented by Kimeswenger et al.~(\cite{kimeswenger}),
Goranskii et al.~(\cite{goranski}), 
Kolev et al.~(\cite{kolev}), Banerjee \& Ashok (\cite{banerjee}),
Wisniewski et al.~(\cite{wisniewski}),
and Crause et al.~(\cite{crause}).
The first spectra obtained after the emersion from solar conjunction
revealed a dramatic temperature decrease of the object, whose
spectrum became dominated by TiO and VO molecular bands, 
suggesting a spectral type later than M10-III 
(Desidera \& Munari \cite{desidera_iauc}).
Noteworthy, a faint blue continuum was found to dominate the spectrum
blueward of 7000~\AA, indicating a likely binary nature for the object.
The hot component was classified as B3V by Munari et al.~(\cite{iau_b3}).
Bond et al.~(\cite{bond03}) studied the light echo using ACS onboard HST.
They found a lower limit to the distance of 6~kpc, implying that V838 Mon
at its maximum brightness was temporarily the brightest star in the Milky 
Way.

In spite of these observational efforts 
the nature of V838 Mon remains largely unknown.

The study of the polarization can shed light on some
physical properties.
On one hand the interstellar polarization gives clues on the 
distance and the absorption toward the object.
On the other, the presence of intrinsic polarization, its wavelength
dependence, its variations during the evolution of the object and 
the polarization across 
line profiles provide clues on the physics of the object.
Wisniewski et al.~(\cite{wisniewski}) present 2-epoch 
spectropolarimetric observations of V838 Mon. They reveal the presence
of intrinsic polarization during the outburst, with variations across
the profile of emission lines.

Here we present the results of our more extensive polarimetric and 
spectro-polarimetric monitoring of V838 Monocerotis, performed at Asiago 
and Crimean observatories from January to November 2002.
The results presented here supersede the premininary analysis of part of 
the same dataset included in  Munari et al.~(\cite{munari02}).


\section{Observations}

\subsection{Asiago}

Polarimetric and spectropolarimetric observations of  
V838 Monocerotis were performed using the
polarimetric mode of AFOSC at the 1.82m telescope at Asiago Observatory
(Italy).
The AFOSC instrument is described by Desidera et al.~(\cite{afosc}).
The polarimeter is presented in detail elsewhere (Pernechele et 
al.~\cite{pol_afosc}) and first results based on its use were presented
by Giro et al.~(\cite{symbiotic}).
Here we recall the main characteristics of the instrument.
The AFOSC polarimeter consists of a double Wollaston prism which 
splits the incoming light into four polarized beams 
(at $0^\circ$, $90^\circ$, $-45^\circ$ and $45^\circ$) 
separated by 20 arcesc. 
These four beams are in principle sufficient to determine the first three 
elements of the Stokes vector, i.e. the intensity $I$ and the two linear 
polarization parameters $Q$ and $U$. The Wollaston can be housed in the filter 
wheel or in the grism wheel of AFOSC. In the first case, spectropolarimetry 
can be performed by inserting a grism in the grism wheel, and in the latter
imaging or photo-polarimetry is obtained by inserting a filter in the
filter wheel.

For the spectropolarimetry, we used three different grisms: 
Grism $\#4$, (wavelength range 4500-7800~\AA; resolution  4.3~\AA/pixel),
Grism $\#7$, (4350-6550 \AA; 2.2~\AA/pixel),
Grism $\#8$, (6250-8000 \AA; 1.8~\AA/pixel).
In all cases a slit 2.5 arcsec wide and 18 arcsec long was used.

Our observational procedure includes spectra taken at position angles
of 0$^\circ$ and 90$^\circ$, to properly eliminate 
the spurious effects introduced by the
different behaviour of the grism for the two polarimetric states.
Flat fields were taken at both slit position angles (0$^\circ$ and 90$^\circ$) 
to avoid spurious polarization effects due to screen reflections.  

Observations were mostly performed in service mode as a target of opportunity,
and this explains some inhomogeneities of the instrument set-up
used and, in some cases, the lack of  observations
on standard stars.

First spectropolarimetric observations were performed very early after 
the first maximum
(January 10), and then we continued our monitoring covering the relevant 
phases of the evolution of this peculiar object for nearly 2 months.
In particular, the maximum of visual magnitude (Feb 2002)
is well covered by our observations.

When the object became too faint for spectropolarimetry, we continued our
monitoring in polarimetric imaging. Deep images obtained in {\em V} band
also allow a study of the polarization of the light echo and of stars 
in the direction of V838 Mon.

Table~\ref{t:spec_obs} presents the journal of observations.

\begin{table}
   \caption[]{Journal of observations. Grism is reported for 
    spectropolarimetric observations ($\#4$, $\#7$, $\#8$, see text for
    details) and filter for polarimetric photometry. A and C refer to
    Asiago and Crimean observations respectively.}
     \label{t:spec_obs}
       \centering

       \begin{tabular}{ccccc}
         \hline
         \noalign{\smallskip}
         Target & Date & UT   &  Grism/Filter &  Obs.   \\
         \noalign{\smallskip}
         \hline
         \noalign{\smallskip}

V838 Mon     & 10/01/2002 & 23 40 & 4                & A  \\ 
V838 Mon     & 11/01/2002 & 23 10 & 4                & A  \\ 
$\beta$ Cas  & 11/01/2002 & 23 10 & 4                & A  \\ 
V838 Mon     & 04/02/2002 & 22 45 & 4,7,8            & A  \\ 
$\beta$ Vir  & 05/02/2002 & 02 10 & 4,7,8            & A  \\ 
HD 93521     & 05/02/2002 & 03 40 & 4,7,8            & A  \\ 
V838 Mon     & 11/02/2002 & 18 40 & {\em U,B,V,R,I}  & C  \\
HD 42807     & 11/02/2002 & 19 30 & 8                & A  \\ 
V838 Mon     & 11/02/2002 & 21 10 & 8                & A  \\
V838 Mon     & 15/02/2002 & 21 00 & {\em U,B,V,R,I}  & C  \\
V838 Mon     & 16/02/2002 & 20 10 & {\em U,B,V,R,I}  & C  \\
V838 Mon     & 18/02/2002 & 21 30 & {\em U,B,V,R,I}  & C  \\
V838 Mon     & 18/02/2002 & 21 50 & 4                & A  \\ 
$\beta$ Vir  & 18/02/2002 & 23 55 & 4                & A  \\ 
V838 Mon     & 04/03/2002 & 21 30 & 4                & A  \\ 
V838 Mon     & 09/03/2002 & 20 00 & {\em U,B,V,R,I}  & C  \\
V838 Mon     & 09/03/2002 & 21 00 & 8                & A  \\
HD 114710    & 09/03/2002 & 02 05 & 8                & A  \\
V838 Mon     & 20/03/2002 & 19 10 & {\em B,V,I}      & A  \\
HD 98421     & 20/03/2002 & 22 20 & {\em B,V,I}      & A  \\ 
HD 204847    & 29/10/2002 & 19 20 & {\em V}          & A \\ 
HD 14069     & 30/10/2002 & 00 25 & {\em V}          & A \\ 
V838 Mon     & 30/10/2002 & 04 20 & {\em V}          & A \\
HD 42807     & 09/11/2002 & 01 55 & {\em V}          & A  \\
HD 43384     & 09/11/2002 & 02 20 & {\em V}          & A  \\ 
V838 Mon     & 09/11/2002 & 03 50 & {\em V}          & A  \\

         \noalign{\smallskip}
         \hline
      \end{tabular}

\end{table}

\subsection{Crimea}

Linear polarimetry of V838 Mon was carried out on  February
11, 15, 16, 18 and March~9, 2002 with the 125cm reflector at the
Crimean astrophysical observatory (Ukraine),  using the computer
controlled  $UBVRI$  Double   Image  Chopping   Photopolarimeter,
developed  at  the  Helsinki  University observatory  by V.~Piirola
(Piirola \cite{piirola73}, \cite{piirola88}). 

The  instrument  has  two   operational  modes:  photometric mode
with elimination of the background  close to the object, and  two
polarimetric   modes   to   measure   linear   and/or    circular
polarization. In  these modes  the background  is measured before
and   after each  set  of  observations  of the object.  An important
advantage is that the sky polarization is directly eliminated  by
using  a  plane  parallel  calcite  plate  as the polarizing beam
splitter.

The   measurements    in   different    colors   are    performed
simultaneously using 5  photomultipliers and dichroic  filters to
split  the  light  into  five  spectral regions centered at 
0.36, 0.44, 0.53, 0.69 and 0.83 $\mu$m with FWHM equal to 0.04,
0.08, 0.08,0.18 and 0.12 $\mu$m respectively. The  instrumental system
is  quite   similar   to   the  Johnson's  {\em UBV}  and Cousins'
{\em R$_{\rm C}$}, {\em I$_{\rm C}$}. 
The  efficiency is high because there  is  practically no
internal absorption in the dichroic beam splitters.

All observations have been carried out in the linear polarization
mode. In the course of polarimetric measurements the
instrument retarder plate rotated with $22.5^{\circ}$ steps in 
the front of  a  polarizer.  One  complete  measurement  consists
of eight integrations  in  different  orientations  of  the 
waveplate. The duration of one measurement with  10 s integration 
time for  each beam  takes  about  3  min.  
All observations  were made  with 10 arcsec  aperture, with the typical
seeing  of  about  3 arcsec.  However,  due  to large zenith angle
($>50^{\circ}$) the photometric quality of the nights was not good.
The  number  of  the  exposures was adapted according to the sky
conditions and the magnitude of the target.


\section{Data Analysis}

\subsection{Asiago}

The reduction of spectropolarimetric data was performed in the standard way 
using IRAF\footnote{IRAF is distributed by 
the National Optical Observatory, which is operated by the Association
of Universities for Research in Astronomy, Inc., under contract with the
National Science Fundation}.
The analysis was performed using our own IDL scripts and the
POLMAP software package maintained by 
Starlink\footnote{www.starlink.rl.ac.uk}. 

A polarization measurement is based on two frames acquired at
slit position angles equal to 0$^\circ$ and 90$^\circ$.
Each image contains four spectra, relative to the four polarization
directions (0$^\circ$, 90$^\circ$, 45$^\circ$ and -45$^\circ$)
produced by the Wollaston prism.
If we label these spectra as $I_{\theta 1}(\lambda)$, $I_{\theta 2}(\lambda)$, 
$I_{\theta 3}(\lambda)$, $I_{\theta 4}(\lambda)$, with 
$\theta$=0$^\circ$,90$^\circ$ the slit position angle,
the two Stokes linear parameters can be computed as:

$$ Q_\theta = \frac{I_{\theta 1}(\lambda)-I_{\theta 2}(\lambda)}
          {I_{\theta 1}(\lambda)+I_{\theta 2}(\lambda)} ~~~~
~~~  U_\theta = \frac{I_{\theta 3}(\lambda)-I_{\theta 4}(\lambda)}
          {I_{\theta 3}(\lambda)+I_{\theta 4}(\lambda)} $$

$$ Q = 	\frac{Q_0 - Q_{90}}{2} ~~~~~~~~ U = 	\frac{U_0 - U_{90}}{2} $$

\begin{figure}
   \includegraphics[width=9cm]{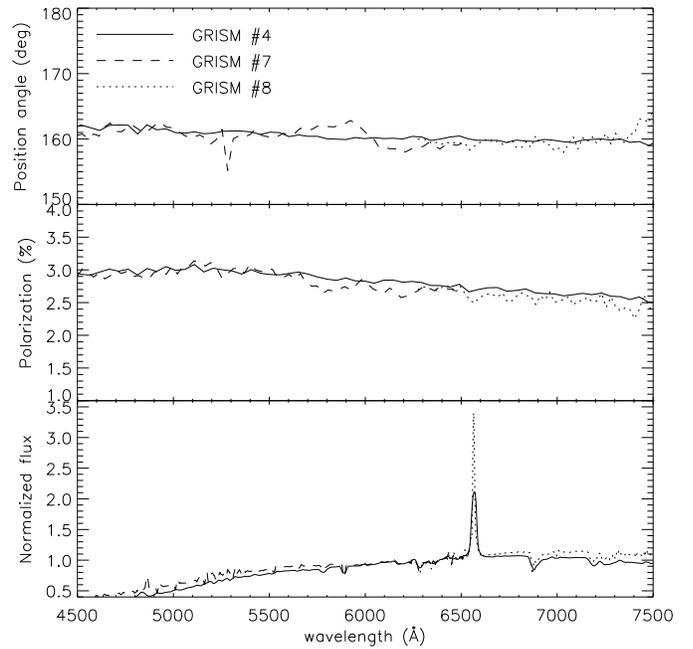}
      \caption{Spectropolarimetry of V838 Mon on Feb 4 2002
               using Grism \#4 (continuous line), Grism \#7 (dashed line) and
               Grism \#8 (dotted line). In the region of overlap, the results
               for the three grism agree quite well. The peak intensity of
               $H_{\alpha}$ emission obviously correlates with spectral
               resolution of different grisms. The feature in position angle
               at $\lambda \sim 5300$~\AA~~is caused by a ghost.}
         \label{f:overlap}
   \end{figure}

On Feb 4, spectropolarimetric observations were obtained 
with all the grisms used in the monitoring.
As shown in Fig.~\ref{f:overlap}, the results of the three grisms in the
overlap region agree quite well. We can therefore safely merge the results
obtained with the different grisms in different nights.

For the polarimetric analysis we used IDL scripts developed
by us for this purpose.
After bias subtraction and flat field division, aperture photometry
is performed for the four polarimetric channels, 
properly subtracting the sky contribution.
The resulting Stokes parameters are calculated as above, but
using only one of the adapter position angles (0$^{\circ}$).

In building the polarimetric map of the light echo (Sec.~\ref{s:lightecho})
the calculation of the Stokes parameters was performed pixel by pixel
in the bidimensional image, after a fine alignment of the four polarimetric
channels. Sky subtraction was performed using POLPACK package.

On some nights zero polarimetric standards were not acquired. However,
the instrumental polarization resulting from the analysis of the 
available standards (including additional observations in dates
different from those of V838 Mon observations) is fairly constant
and always below 0.3\%. Continuous observations of the same standard
at different telescope positions and airmasses show a dispersion
at 0.2\% level, mainly due to the scattering of light at slit edges
(Lorenzi \cite{tesi_lorenzi}).
The analysis of standard stars with known high polarization 
indicates that systematic errors
in the position angle are below 2.0$^{\circ}$. 

\subsection{Crimea}

Before  reduction  all  raw  data  are  checked  to eliminate bad
points from the observations. The data reduction was carried  out
using the software package  developed by V.~Piirola, which  performs
all the required corrections (background subtraction,  correction
for  the  instrumental  polarization,  transformation  into   the
equatorial  system,  transformation  from instrumental magnitudes
to  the  standard  system,  calculation  of  the weighted nightly
means  of  the  normalized  Stokes  parameters  of   polarization).

Polarimetric    observations   are   corrected   for  instrumental
polarization by applying the instrumental  constants calculated for a
number  of polarimetric standards. 
The data were reduced with instrumental constants for the time
interval of our observations (Feb-Mar 2002).
Unfortunately,  due  to  a problem
during   the  realuminization  process   of  the  main  mirror, the
instrumental  polarization  is  large  and  wavelength  dependent
decreasing from 2.4\% in {\em U} to 0.5\% in {\em I} band.
An inspection of the instrumental constants estimations obtained 
before and after the observations of V838~Mon (January-July 2002)
has shown that their scattering does not exceed 0.1\% in {\em U}
and is at the level of 0.02\% in the other bands.


\section{Results}

\subsection{The polarization of V838Mon}
\label{s:pol}

The results of our measurement of polarization of V838 Mon are summarized
in Table \ref{t:pol_evolution}.
To allow a comparison between  the spectro-polarimetric
results and the polarimetric photometry, the integrated 
{\em V} band polarization was measured on the spectra. Quoted errors
of Asiago observations are those due to photon noise.
Errors of Crimean data include photon noise and
the error due to the scattering of the Stokes parameters
between all 8 integrations  of a complete single measurement.
The Stokes parameters are weighted by the largest of these errors. The
errors of  the  final data are calculated by the usual formula for
weighted error.

The {\em V} band polarization ranges between 2.35 and 3.09\%.
The results of Asiago and Crimea observations agree
fairly well, with differences at the 0.1\% level.
The wavelength range and effective wavelength of
photometric bands depend on the spectral type of the object
and may be significantly different from the standard ones
for very red objects like V838 Mon.
This could explain some systematic differences (i.e.~polarization in
{\em R$_{\rm C}$} and {\em I$_{\rm C}$} bands lower 
than the spectropolarimetric data, see Fig.~\ref{f:russi}).
There is instead a  position angle offset between the
two datasets larger than the quoted  errors.
The position angle is about 159$^{\circ}$ and 153$^{\circ}$ for the
Asiago and Crimea data, respectively. 

\begin{figure}
   \includegraphics[width=9cm]{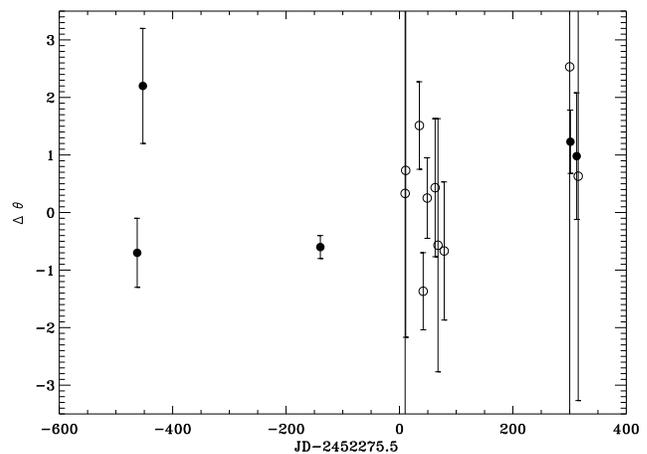}
      \caption{Stability of position angle measurements with AFOSC
               polarimeter. Filled circles: difference between position
               angle of high-polarization standard stars measured by us
               and literature values. Empty circles: position angle
               measurement of V838 Mon taken with AFOSC polarimeter 
               (with the weighted average 159.1$^{\circ}$ subtracted). 
               Interstellar 
               polarization dominates at all epochs, so that position angle
               of V838 Mon is expected to be nearly constant.
               The time coordinate is the same of Fig.~\ref{f:pol_evolution}.
               The results of standard stars leaves little room for 
               systematic errors in position angle measurement with AFOSC
               polarimeter exceeding 2$^{\circ}$.}
         \label{f:check_standard}
   \end{figure}

We investigated in some detail the absolute calibration
of position angle.
Polarized standard stars were not observed
at Asiago in January-March 2002; however the analysis
of such standards taken at different epochs (including
October and November 2002) reproduce the known position
angles within 2$^{\circ}$ (Fig.~\ref{f:check_standard}).
A similar analysis performed on Crimean polarimeter  shows that
typical systematic errors on position angles do not exceed
 1$^{\circ}$.5.
The discrepancy in position angle between Asiago and Crimea observations
is larger than the corresponding uncertainties. In spite of our best efforts
put in investigating the matter, no explanation this has been obtained.
However, this fact does not significantly affect the main results
of this study, since the absolute values of polarization
agree {\it very} well.

The dependence of the polarization on wavelength is smooth and 
compatible by a Serkowski law (Serkowski et al.~\cite{Serkowski75})
with $\lambda_{max}$ between $4928\pm21$~\AA\ (Feb 4)
and $5751\pm52$~\AA\ (Feb 18). 
Position angle is fairly constant in the observed wavelength range
for the Asiago spectropolarimetry while a mild trend with wavelength is
present in the Crimea data.

\begin{table*}
   \caption[]{Polarization of V838 Mon. The integrated {\em V} band 
              polarization was computed from the spectra to allow
              the study of the polarimetric evolution of the object.
              A and C refer to Asiago and Crimea observations respectively.}
     \label{t:pol_evolution}
       \centering

  {\scriptsize

       \begin{tabular}{ccccccccccc}

         \hline
         \noalign{\smallskip}
         Date  &  $P_{U}$ & $\theta_{U}$ &  $P_{B}$ & $\theta_{B}$ &  $P_{V}$ & $\theta_{V}$ &  $P_{R}$ & $\theta_{R}$ &  $P_{I}$ & $\theta_{I}$  \\

         \noalign{\smallskip}
         \hline
         \noalign{\smallskip}

10/01/02 A & & & & & 2.74$\pm$0.59 & 159.4$\pm$7.0 & & & & \\
11/01/02 A & & & & & 2.61$\pm$0.29 & 159.8$\pm$2.9 & & & & \\
04/02/02 A & & & & & 2.92$\pm$0.10 & 160.6$\pm$0.8 & & & & \\
11/02/02 A & & & & & 2.80$\pm$0.06 & 157.7$\pm$0.7 & & & & \\
11/02/02 C & 2.30$\pm$0.36 & 154.6$\pm$4.5 &2.68$\pm$0.10& 153.7$\pm$1.1 & 2.66$\pm$0.06 & 155.1$\pm$0.7 & 2.56$\pm$0.04 & 154.7$\pm$0.5 & 2.39$\pm$0.04 & 154.6$\pm$0.5\\
15/02/02 C & 2.12$\pm$0.17 & 148.0$\pm$2.3 &2.41$\pm$0.04& 151.8$\pm$0.5 & 2.36$\pm$0.04 & 152.3$\pm$0.5 & 2.28$\pm$0.04 & 153.3$\pm$0.4 & 2.09$\pm$0.03 & 153.0$\pm$0.5\\
16/02/02 C & 2.29$\pm$0.17 & 150.2$\pm$2.2 &2.50$\pm$0.06& 152.3$\pm$0.7 & 2.47$\pm$0.04 & 152.7$\pm$0.5 & 2.33$\pm$0.02 & 153.0$\pm$0.3 & 2.09$\pm$0.02 & 152.8$\pm$0.2\\
18/02/02 A & & & & & 2.49$\pm$0.05 & 159.3$\pm$0.7 & & & & \\
18/02/02 C & 2.43$\pm$0.29 & 145.5$\pm$3.4 &2.30$\pm$0.06& 151.1$\pm$0.7 & 2.43$\pm$0.04 & 152.9$\pm$0.4 & 2.31$\pm$0.02 & 154.4$\pm$0.3 & 2.11$\pm$0.02 & 153.3$\pm$0.2\\
04/03/02 A & & & & & 2.58$\pm$0.10 & 159.5$\pm$1.2 & & & & \\
09/03/02 A & & & & & 2.54$\pm$0.18 & 158.5$\pm$2.2 & & & & \\
09/03/02 C & 2.32$\pm$0.23 & 150.7$\pm$2.8 &2.53$\pm$0.06& 152.7$\pm$0.7 & 2.52$\pm$0.04 & 153.1$\pm$0.4 & 2.30$\pm$0.02 & 153.9$\pm$0.3 & 2.14$\pm$0.02 & 153.5$\pm$0.3\\
20/03/02 A & & &2.40$\pm$0.09&160.1$\pm$1.6& 2.35$\pm$0.07 & 158.4$\pm$1.2 & & & 2.07$\pm$0.09 & 157.2$\pm$2.1\\
28/10/02 A & & & & & 3.09$\pm$0.58 & 161.6$\pm$6.4 & & & & \\
12/11/02 A & & & & & 3.05$\pm$0.35 & 159.7$\pm$3.9 & & & & \\

         \noalign{\smallskip}
         \hline
      \end{tabular}
}
\end{table*}

\subsection{Time Variability}

\begin{figure}
   \includegraphics[width=9cm]{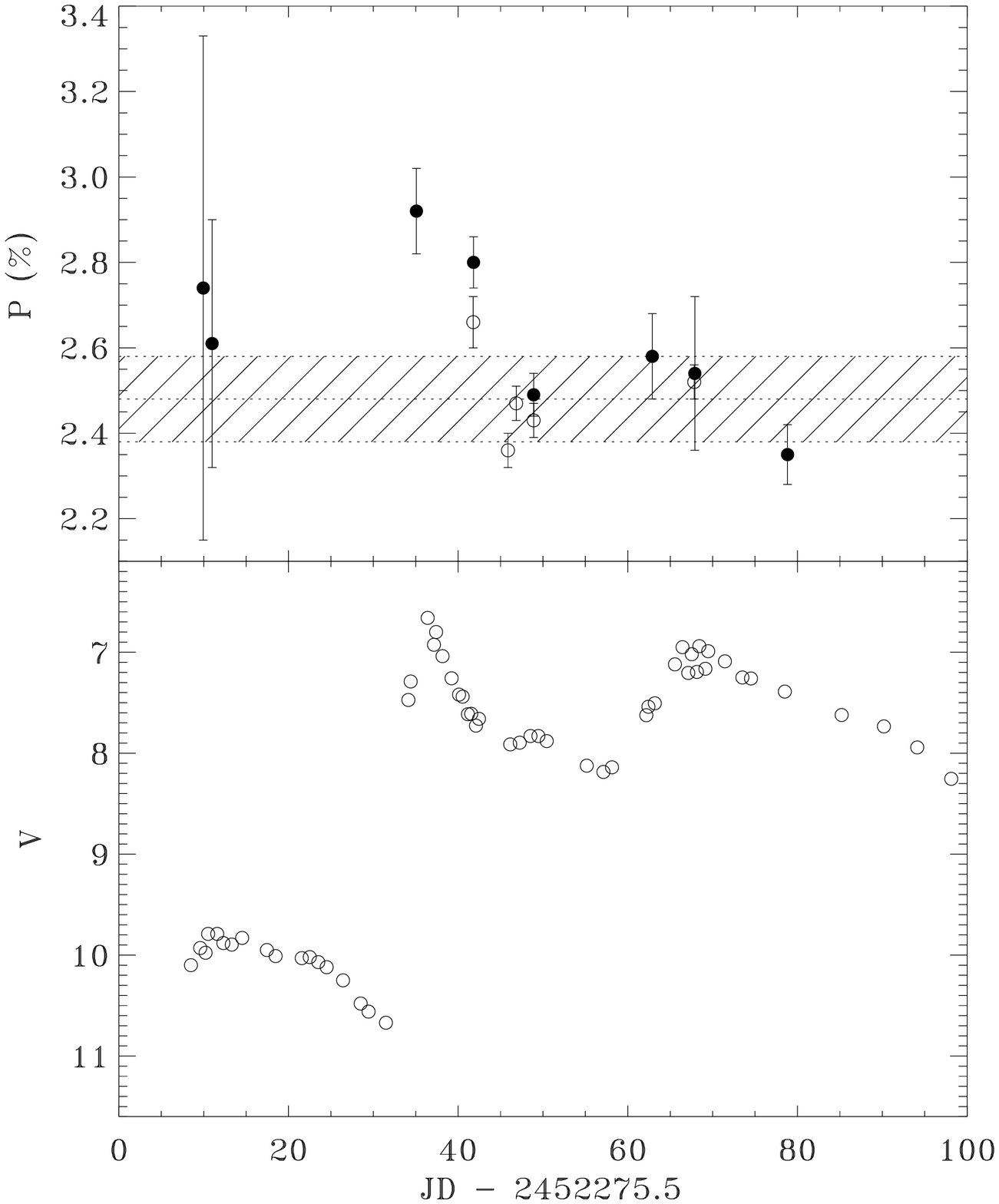}
      \caption{Evolution of {\em V} band polarization  
               of V838 Monocerotis during the outburst (upper panel). 
               Filled and empty circles represent
               the Asiago and Crimea data respectively. The shaded area
               shows our estimate of interstellar polarization
               ($2.48\pm0.10$\% in {\em V} band).
        The lower panel shows the photometric evolution of the object
        (from  Munari et al.~\cite{munari02}). The time coordinate 
        is that used  by Munari et al.~(\cite{munari02}). The
        intrinsic polarization appears correlated with the visual
        magnitude during the second and possibly the third maxima
        of the object. Two further polarization measurements at later epochs
        are outside of the plot limits.}
         \label{f:pol_evolution}
   \end{figure}

The variations of the polarization during the outburst are small but 
significant and they appear to have some connection with the 
photometric behaviour of the object (Fig.~\ref{f:pol_evolution}). 
The maximum of polarization during the outburst (2.92\% in {\em V} band) 
was measured on Feb~4, two days after the beginning of the rapid rise
of luminosity of the object and two days before the visual maximum.
One week later, the polarization
was $\sim 0.1$\% lower while after a further week it stabilized
at $\sim 2.5\%$, 
with a possible increase by $\sim 0.1-0.2$\% during 
the third peak of the {\em V} band light curve (March 2002).
The polarization in {\em V} band in October-November 2002 (not shown in
Fig.~\ref{f:pol_evolution}) possibly increases back
to higher values ($\sim3.1$\%) but the errors are larger due to the 
much fainter magnitude of the object (V=16.05, Munari et al.~\cite{iau_b3})
and therefore we do dot speculate any further about it. 
The polarization contamination due to the patchy nature of the 
light echo is estimated to be smaller than the photon noise.

We note that the {\em V} band flux in October-November 2002 
is dominated by the hot B3 component
(Munari et al.~\cite{iau_b3}), with the cool component contributing 
$\sim 10$\% of the light at most.

\subsection{The Interstellar Polarization}
\label{s:interstellar}

The small time variability of the polarization of V838 Mon during and
after the outburst, the constancy of the position angle, the polarization 
in the core of $H_{\alpha}$ and the wavelength dependence of the 
polarization, compatible with the Serkowski law, 
indicate that most of the observed polarization is of interstellar origin
(Mc Lean \& Clarke \cite{mclean79b}).

\begin{figure}
   \includegraphics[width=9cm]{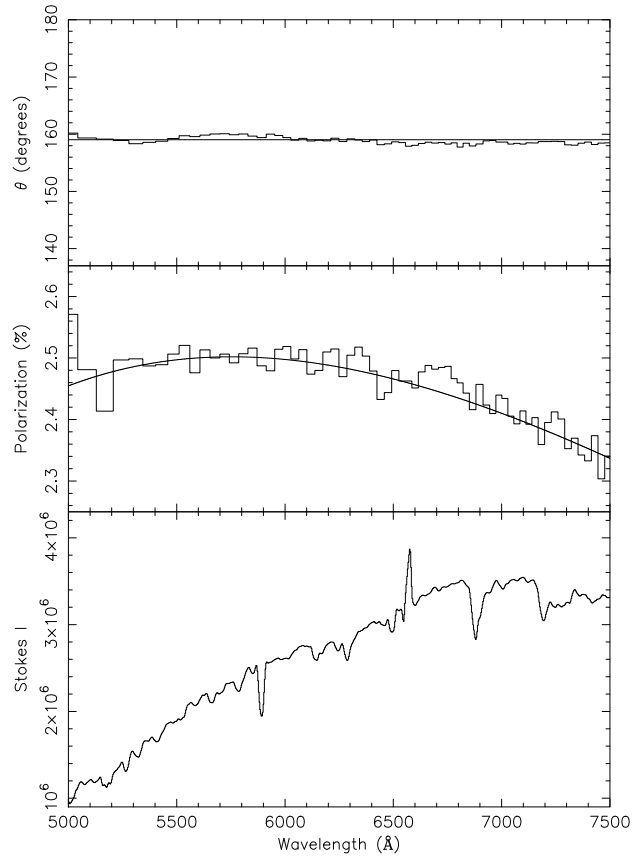}
      \caption{Position angle, polarization and intensity on 18th Feb 2002, 
               mainly showing interstellar
               polarization. The fit of the Serkowski law as modified by
               Whittet et al.~(\cite{whittet}) is overplotted.}
         \label{f:Serkowski}
   \end{figure}

\begin{figure}
   \includegraphics[width=9cm]{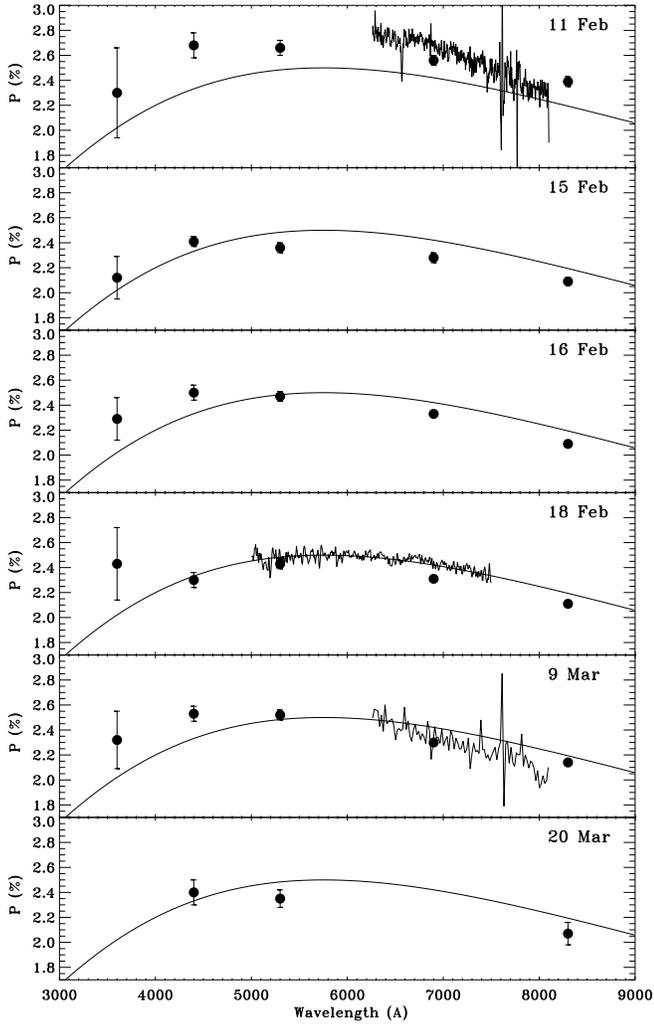}
      \caption{Polarimetric photometry from Feb~11 to Mar~20 2002.
               The spectro-polarimetric data taken on Feb 11,
               Feb 18 and Mar 9 are overplotted on the corresponding
               panel. 
               The fit of the Serkowski law performed on the spectrum
               of Feb 18 is overplotted on each panel. 
               The 11th Feb data are clearly
               above the fit, indicating the presence of intrinsic
               polarization, demonstrated also by the depolarization
               of $H_{\alpha}$ emission line (see below). 
               The other data are in 
               reasonable agreement, indicating that after Feb 15
               the polarization is mainly interstellar.
               Some discrepancy between broad band  and low resolution
               spectroscopic polarimetry may be due to the very
               peculiar spectral energy distribution of the object
               ($B-V > 1.5$ at the epochs of polarimetric photometry).}
         \label{f:russi}
   \end{figure}

To estimate the value of the interstellar polarization we consider
the following points:
\begin{itemize}
\item
the polarization is fairly constant after Feb~15 2002, with 
a possible marginal increase (0.1\%) during the third peak
of the light curve;
\item
the fit of the Serkowski law as modified by Whittet et al.~(\cite{whittet}) 
to the Feb 18 spectrum gives $P_{max}=2.50\pm0.05 $, 
{\bf $\theta_{max}=159.0^{\circ}$} at 
$\lambda_{max}=5751\pm52$~\AA~~(Fig.~\ref{f:Serkowski}). 
Such a value of $\lambda_{max}$ is typical for the interstellar medium;
\item
the residual polarization in the core of $H_{\alpha}$ on 11th Feb spectrum 
is 2.35\% with $\theta=159^{\circ}$. 
Using the Serkowski law derived on Feb~18
this gives $P_{max}=2.46\pm0.06$ at 5751~\AA, fully compatible with
the previous estimate.
\end{itemize}

From these fully consistent estimates we adopt as interstellar
polarization $P_{max}=2.48\pm0.10$ (including zero point 
uncertainty) 
with the wavelength dependence obtained from the Feb~18 spectrum
($\lambda_{max}=5751\pm52$~\AA)\footnote{The fit of the 
Serkowski law as modified by 
Whittet et al.~(\cite{whittet}) to the polarimetric data
is fully consistent with that from spectropolarimetry concerning
the value of polarization ($P_{max}=2.49\%$) while the 
$\lambda_{max}$ is bluer by about 500~\AA. In spite of the wider
wavelength coverage, the wavelength dependence of the interstellar
polarization from photometric data is much more uncertain because
of the limited number of bands (one of which affacted by 
large internal errors) and  
the uncertainty in the effective 
wavelengths for an object with such peculiar  spectral energy 
distribution as V838~Mon.}

Considering the mentioned systematic discrepancy between Asiago and
Crimea data, we adopt as position angle  $\theta=156 \pm 3^{\circ}$.
In the subtraction of the interstellar polarization from Asiago spectra,
we consider instead the position angle of Asiago dataset 
($\theta=159^{\circ}$) for consistency.

The contribution of the
circumstellar material not directly associated with the outburst, whose
presence was evidenced by the light echo and the IRAS detection of the
progenitor (Munari et al.~\cite{munari02}, 
Kimeswenger et al.~\cite{kimeswenger}), could add to the fraction
of purely interstellar polarization.
Hereafter we will refer to interstellar polarization 
as also including  such (possible) circumstellar effects.

\subsection{Intrinsic Polarization}
\label{s:intrinsic}

\begin{figure}
   \includegraphics[width=9cm]{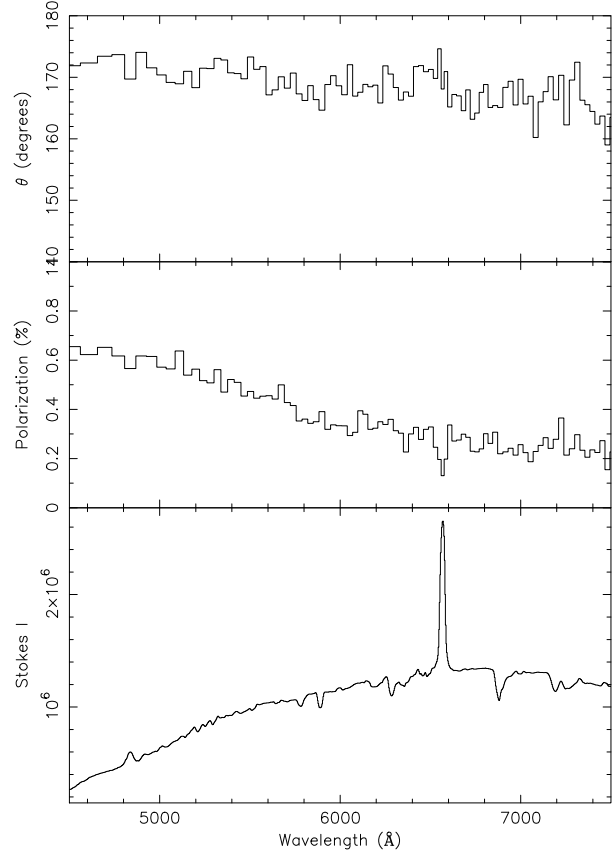}
      \caption{Intrinsic polarization on Feb 4 2002. The insterstellar
               polarization, taken from the Serkowski fit of
               Feb 18 2002 (see Fig.~\ref{f:Serkowski}), was subtracted.
               Note the increasing polarization toward shorter wavelengths
               and the possible partial depolarization of $H_{\alpha}$.}
         \label{f:intrinsic}
   \end{figure}

To make more evident the wavelength dependence and the time variability of the 
intrinsic polarization, we subtracted the interstellar polarization 
derived in Sec.~\ref{s:interstellar} from the observed one 
(Fig.~\ref{f:intrinsic}).
The maximum of {\it intrinsic} polarization is reached on the Feb 4 spectrum.
The intrinsic polarization shows a marked increase toward shorter 
wavelength, reaching 0.7\% at 5000~\AA, with position angle 
$\theta=170^{\circ} $.
Polarimetric data on 11 Feb also show the presence of intrinsic
polarization, about 0.3\%, with position angle $175\pm6^{\circ}$.
The wavelength dependence of the intrinsic polarization indicates
that further mechanism(s) beside Thomson scattering are at work.

A trend of intrinsic polarization increasing toward shorter wavelength 
in the range 4500-7500~\AA~has been observed in several other objects.
In the case of the Be star $\zeta$ Tau a
complex pattern of polarization as a function of wavelength was observed, with
large discontinuities at the Balmer and Paschen edges and a polarization
increasing blueward in the Paschen continuum (Wood et al.~\cite{wood97}).
This can be explained in terms of electron scattering modified
by continuous hydrogen absorptive opacity (Wood et al.~\cite{wood96}).
However, our data do not reach the Paschen and Balmer jumps, so that
we cannot conclude that the observed trend is due to this effect. 
A possible Paschen jump is seen in the Wisniewski et al.~(\cite{wisniewski})
data, but only at the one-sigma level.

The observed trend could be ascribed also to a dust disk formed
in the stellar outflow (see e.g. Clayton et al.~\cite{clayton97}).

In the case of nova V4444 Sgr (Kawabata et al.~\cite{kawabata00}),
the variation of polarization with wavelength was instead
interpreted as Mie scattering due to 
a preexisting circumstellar dust cloud at fairly large
distance from the object. This explanation is probably not correct
for V838 Mon since the intrinsic polarization is variable on fairly short
timescales. 

\subsection{Polarization across spectral lines}
\label{s:lines}

The resolution of the spectra obtained with Grism \#8
enables study of the polarization across spectral lines.  For example,
V838 Mon developed a prominent $H_{\alpha}$ emission line during the second
maximum.

\begin{figure}
 \vspace{0.3cm}
  \includegraphics[width=8.8cm]{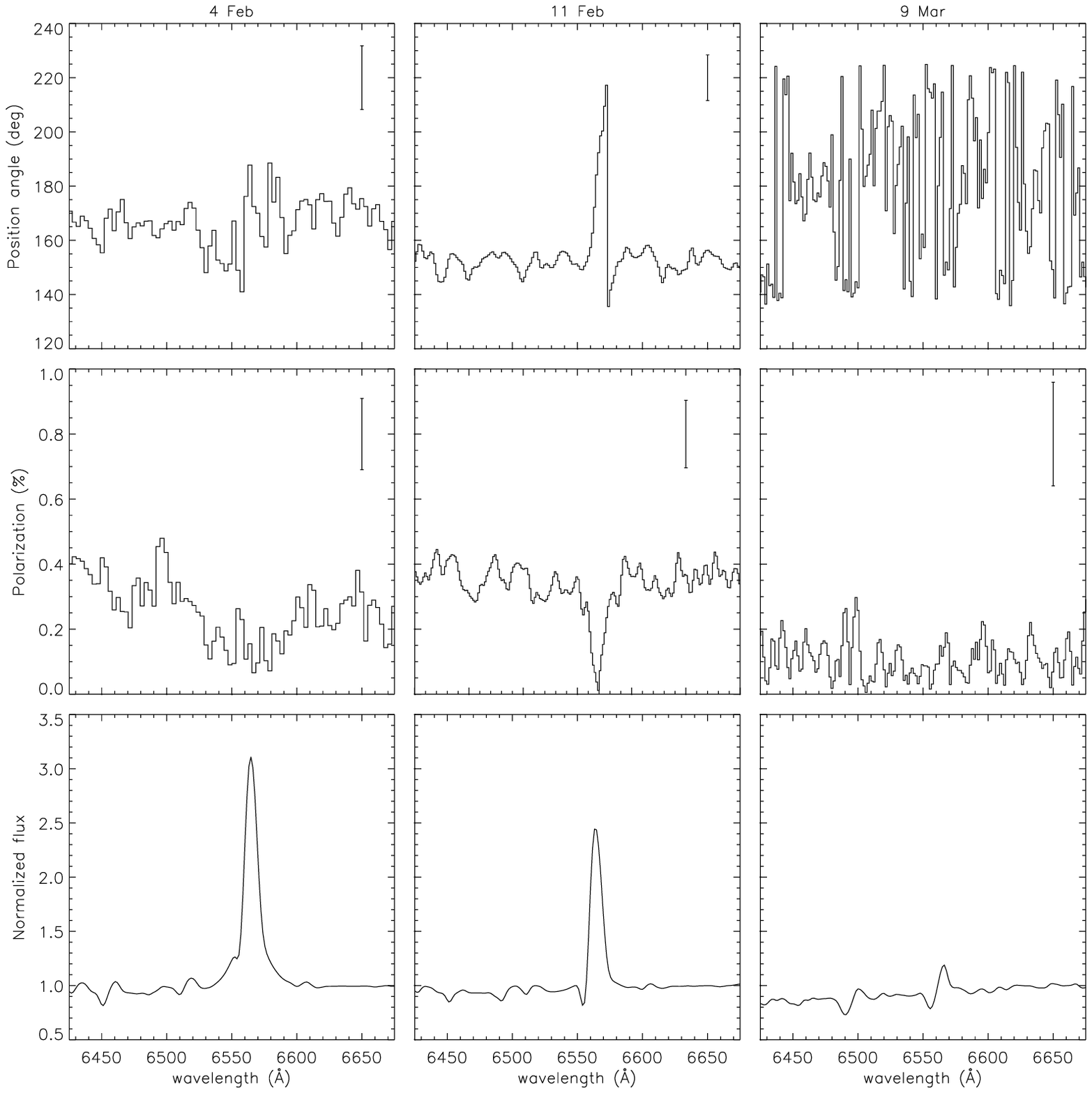}
   \vspace{0.3cm}
      \caption{Polarization across $H_{\alpha}$ for the spectra
               obtained with Grism \#8 (from top to bottom) on Feb 4,
               Feb~11 and March~9 2002. The interstellar polarization,
               estimated as the polarization in the core of $H_{\alpha}$
               on the 11th Feb spectrum, was subtracted.
               Typical errorbars of polarization and position angle 
               are overplotted. 
               The very noisy position angle on 9th March spectrum is
               the result of the very low intrinsic polarization.}
         \label{f:halpha}
   \end{figure}

Fig.~\ref{f:halpha} shows the polarization around $H_{\alpha}$
at three epochs (subtracting the interstellar polarization).
A marked depolarization, at $\sim 0.4$\% level, is evident in the
spectrum of Feb~11, 5 days after the maximum of the {\em V} light curve.
A change in the position angle across the line profile is also clearly seen.

A marginal depolarization of $H_{\alpha}$, possibly characterized
by a broader profile, may be present
in the Feb~4 spectrum  (two days before
the peak of the light curve),  when $H_{\alpha}$ was even brighter
but was characterized by 
a very different profile, without the P-Cyg absorption and very 
broad wings.
No depolarization was detected 
on Mar~9, when $H_{\alpha}$ emission was significantly fainter
and the intrinsic polarization of the object was small 
($\sim 0.1-0.2$\% at most). 
The result of Wisniewski et al.~(\cite{wisniewski}) indicate complete
$H_{\alpha}$ depolarization on 8th Feb.


\subsection{Comparison with Wisniewski et al.~(\cite{wisniewski})}
\label{s:other}

Wisniewski et al.~(\cite{wisniewski}) presented spectropolarimetry 
of V838 Mon obtained in two epochs, Feb 8 and Feb 13.
On Feb 8 the {\em R}-band polarization was 3.2\%. 
The depolarization across $H_{\alpha}$ line at this date 
and the lower polarization of their later spectrum 
(P=2.67\%) indicate the presence of intrinsic polarization.

Our data on polarization of V838 Mon agree qualitatively with 
these results. 
Our measurements do not reach polarization as high as 3.2\%
during the outburst, but this may 
be due to intrinsic variability of the object
and different observing dates. We noted 
that the polarization seems to follow the photometric evolution
(Fig.~\ref{f:pol_evolution}), so that the maximum
could have occurred between our observations of Feb~4 and 11 2002. 
The polarization measured by Wisniewski et al.~(\cite{wisniewski}) 
on Feb~13 is intermediate between ours on Feb 11 and 18. 
Their position angle is 153$^{\circ}$, similar to that resulting
from Crimea data (see Sect~\ref{s:pol}).

We note that our data on Feb~11 reveal the presence of 
intrinsic polarization at the $\sim0.3-0.4\%$ level, so that the assumption 
by Wisniewski et al.~(\cite{wisniewski}) that the polarization on 13th Feb
was purely interstellar is questionable.
Furthermore, a visual inspection of their Fig.~4 reveals that the
minimum of polarization in the core of $H_{\alpha}$ is 
lower (by $\sim 0.2$\%) than their fit for the interstellar 
polarization and more compatible
with our estimate.

The trend of intrinsic polarization with wavelength of their Feb~8
spectrum is remarkably similar to our on Feb~4 in the overlap region.
However, their position angle for the intrinsic polarization 
($\theta=127^{\circ}$) is quite different from our determination.

\subsection{The polarization of stars in the field of V838 Mon}
\label{s:pol_field}

The study of the polarization of field stars is useful to confirm
our estimate of interstellar polarization and it 
constrains the distance and reddening estimates of the target.

Towards this aim, we measured the polarization of about 40 field stars
in the direction of V838 Mon on our deep polarimetric {\em V} 
images taken on October and November 2002.
Stars projected inside the light echo were not considered. 

Errors are quite large due to the faintness of most of the measured stars; 
however it appears that about half of the field stars with polarization
larger than 0.5\%
have position angles similar to that of the estimated interstellar polarization
of V838 Mon. This is expected for field stars seen toward coherent
portions of interstellar medium.
Furthermore most of the stars with such position angle have lower
polarization than V838 Mon.
All these stars are within 5 arcmin from V838 Mon, so that the 
spatial inhomogeneities of the interstellar medium should likely
play a marginal role. Therefore we can infer that V838 Mon 
is at larger distance from us than most of the surrounding field stars.
However, the lack of distance estimates for these stars
does not allow us to quantify this result in terms of a lower limit to
the distance.
Observational efforts aimed at obtaining their distances 
using spectral classification are in progress.
Details will be presented elsewhere.

\subsection{The Polarization of the Light Echo}
\label{s:lightecho}

\begin{figure}
   \includegraphics[width=5cm]{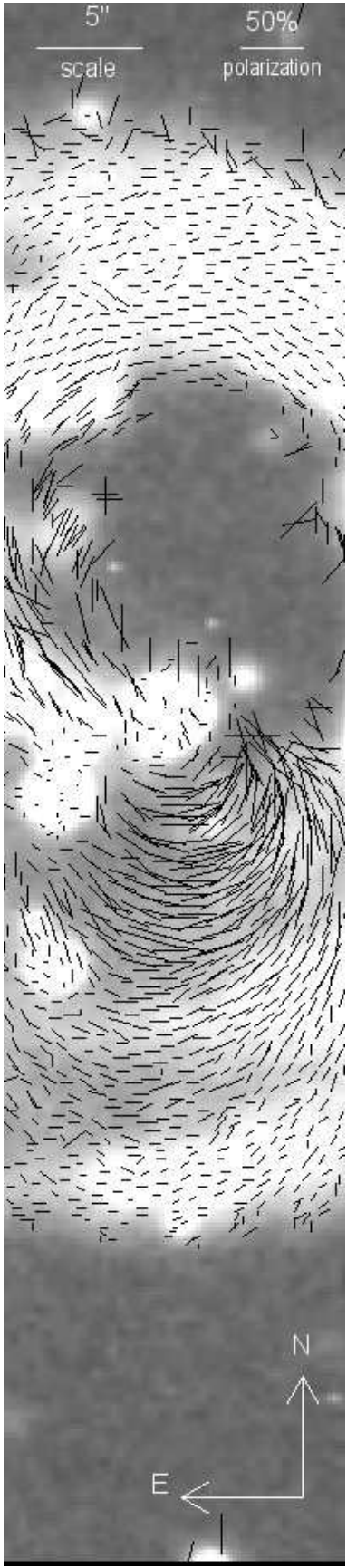}
      \caption{Polarization map of a portion of 
               the light echo of V838 Monocerotis on Nov 9 2002.
               The central object is V838 Mon.
Typical errors on polarization span from 1.5\% in the brightest 
zones  to 5\%.}
         \label{f:lightecho}
   \end{figure}

Our deep polarimetric imaging of the field of V838 Mon 
(Nov 9 2002)
covers the whole extension of the light echo in the
north-south direction and 18 arcsec east-west (Fig.~\ref{f:lightecho}).  

The light echo shows a distributed polarization with
a complex pattern, 
reaching a 45\% maximum and $88^{\circ}$
position angle at 11 arcsec south and 3 arcsec west of the central star. 
Typical errors on polarization span from 1.5\% in the brightest 
zones  to 5\% on the faintest.
Note that the position angle quoted by Giro et al.~(\cite{lightecho_iauc})
is wrong by $90^{\circ}$. 

A large polarization of the light echo was
also reported by Bond et al.~(\cite{bond03}).
On their HST polarimetric images (not  fully calibrated), obtained
two months earlier than ours (when the light echo appearence was understandably
different) they obtained a peak polarization ``of 50\% at the inner rim of
the cavity, south east of the star''. The {\em very} different spatial
resolution between ground-based and HST observations also plays a role,
in addition to changing aspect of the light echo with time, in accounting for
the differences between our and Bond et al. location for peak polarization
while maximum values however fairly well agree. Bond et al.~(\cite{bond02b})
observations on Apr 30 also indicate that the polarization degree
and the position of the maximum evolves with time.

The general appearance is that  of a dipole with its axis
roughly oriented in the east-west direction and  a
more visible southern lobe. Such a structure is  different
than expected from light scattered by an homogeneous
circumstellar medium (see, e.g. Sparks \cite{sparks}).
We also note that about 14 arcsec south of the central object,
the polarization degree of the lobe suddenly drops from $\sim 40\%$ down
to $\sim 17\%$.
The northern lobe might possess quite a similar structure
as suggested by the fainter features observed close to the star.
Due to the limitation of our field of view in the east-west direction,
we are unable to assess the geometrical structure of the
polarization in the external regions. However, the general constancy
of the polarization degree is quite evident.

\section{Clues on reddening and distance}
\label{s:red_dist}

The distance to V838 Mon is very uncertain, with literature estimates 
ranging from 650 pc (Kimeswenger et al.~\cite{kimeswenger}) to 10 kpc
(Munari et al.~\cite{iau_b3}).
Reddening is likely constrained in the range {\em E(B-V)}$ \sim$  0.4-0.9.
A further estimate of  the reddening can be derived 
using the limit  {\em E(B-V)} $> P_{interstellar}/9$ 
(Serkowski et al.~\cite{Serkowski75}). 
This gives {\em E(B-V)} $> 0.28$.
Adopting the more typical value 
{\em E(B-V)} $\sim P_{interstellar}/5$ we get {\em E(B-V)} $\sim 0.50$.
This is similar to the value {\em E(B-V)}=0.50 adopted by
Munari et al.~(\cite{munari02}) and to {\em E(B-V)}=0.54 recently found by
Wagner \& Starrfield (\cite{wagner}) from the equivalent width
of Na I D lines in the spectrum of the blue component of the
system. It is instead smaller than {\em E(B-V)}=0.80 
quoted by Zwitter \& Munari 
(\cite{zwitter}) based on the intensity of Na I and K I interstellar lines
on outburst spectra,
{\em E(B-V)}=0.7 derived by Kimeswenger et al.~(\cite{kimeswenger}) 
by photometric and spectroscopic observations during of the outburst and 
{\em E(B-V)}=0.9 estimated by Munari et al.~(\cite{iau_b3}) from 
spectral classification and photometry of the hot component.
The larger values of reddening, if confirmed, would indicate 
a ratio  $P_{interstellar}/${\em E(B-V)} $\sim 3$, that is still
fairly normal (see Serkowski et al.~\cite{Serkowski75}).

The wavelength dependence of the interstellar polarization, 
following a Serkowski law with $\lambda_{max}=5751 \pm 52$~\AA, implies
a normal ratio of total to selective absorption ($R_{V}=3.22 \pm 0.17$ 
using the calibration by Whittet \& van Breda \cite{whittet78})\footnote{This
is also fully consistent with the dependence of $R_{V}$ on galactic latitude
derived by Whittet (\cite{whittet77}).}.
This leaves little room for the presence of a significant 
circumstellar absorption characterized by peculiar dust--to--gas ratio
or grain composition and nature.
The similar polarization angle between V838 Mon and a large fraction 
of field stars also indicates a minor contribution of the circumstellar 
material to the observed reddening.

The polarization larger than most of the field stars and 
the value of reddening likely in the range 0.50-0.80
point to a fairly large distance, considering that 
on the reddening maps by Nekel \& Klare (\cite{nekel})
such values of reddening are reached  for distances
larger than 3 kpc. This is in qualitative agreement 
with the recent results by Bond et al.~(\cite{bond03}).

\section{Conclusion}

We have monitored the polarization of V838 Mon from January to November 2002
covering the relevant phases of the evolution of this mysterious object.
The main results of this study are:
\begin{itemize}
\item
The observed polarization of V838 Mon varies between 2.4 and 3.1\%.
\item
Interstellar (+ circumstellar) polarization  is $\sim 2.48$\% 
with position angle $\theta=156\pm3^{\circ}$.
This represents the major contribition to the observed value.
\item
The intrinsic polarization seems to follow the light curve 
of V838 Mon, reaching a maximum of 0.7\% at 5000~\AA\ on Feb 4,
and then quickly declining to zero in $\sim$ 10 days.
The occurrence of intrinsic polarization could be explained by departures
from spherical geometry during the outburst.
\item
The intrinsic polarization of Feb 4 shows
a marked wavelength dependence, increasing toward shorter wavelengths.
This trend is compatible with that expected in case of 
electron scattering modified
by continuous hydrogen absorptive opacity (Wood et al.~\cite{wood96}),
but our limited wavelength coverage does not allow conclusive results
to be reached. 
\item
$H_{\alpha}$ emission line is depolarized in the spectrum of Feb 11.
\item
The analysis of the polarimetric map of the light echo indicates
a complex polarization pattern, reaching $P_{max}=45$\%.
\item
The interstellar polarization implies  {\em E(B-V)} $> 0.28$.
A higher reddening {\em E(B-V)} $\sim 0.5$, 
in better agreement with independent estimates, is obtained
adopting a more typical value for the ratio  $P_{interstellar}/${\em E(B-V)}.
\item
A normal ratio of total to selective absorption ($R_{V}=3.22\pm0.17$) 
is derived from the wavelength of maximum interstellar polarization,
suggesting a normal interstellar medium toward V838 Mon
\end{itemize}





\begin{acknowledgements}

   We warmly thank the Asiago Observatory Staff for its support
   during observations and G.~Baume for observing V838 Mon for us during
   his observing time.
   We are  grateful  to  V.~Piirola  for  the  permission  to  use  his
   photopolarimeter in these observations.
   We thank the referee Dr. G. Clayton for its stimulating comments.
   This research made use of the 
   SIMBAD database, operated at CDS, Strasbourg, France.
   TT acknowledges support by Polish KBN Grant No. 5 P03D 003 20.

\end{acknowledgements}

\end{document}